\documentclass[letterpaper,twocolumn,preprintnumbers,secnumarabic,amsmath,amssymb,superscriptaddress]{revtex4}
\usepackage{amssymb,amsmath}
\usepackage{ulem} 
\usepackage{bm} 
\usepackage{booktabs} 
\usepackage{array}
\usepackage{latexsym}
\usepackage{graphicx}
\usepackage{color}
\usepackage{datetime}

\usepackage{verbatim}
\usepackage{chngpage} 

\usepackage{psfrag}

\usepackage{mciteplus}

\usepackage[colorlinks=true,      linkcolor=blue,      urlcolor=blue,      
            filecolor=blue,      citecolor=magenta,       pdfstartview=FitH,     
                                                pdfpagemode=UseNone,      bookmarksopen=true]{hyperref}  
\usepackage[all]{hypcap}     

\begin{document}

\title{Holographic correlators with BPS bound states in ${\mathcal N}=4$ SYM}

\author{Francesco Aprile}
\affiliation{Departamento de F\'\i sica Te\'orica \& IPARCOS,  Facultad de Ciencias F\'\i sicas, Universidad Complutense de Madrid, 28040 Madrid, Spain}

\author{Stefano Giusto}
\affiliation{Dipartimento di Fisica,  Universit\`a di Genova and I.N.F.N. Sezione di Genova, Via Dodecaneso 33, 16146, Genoa, Italy}

\author{Rodolfo Russo}
\affiliation{School of Mathematical Sciences,
 Queen Mary University of London, Mile End Road, London, E1 4NS,
United Kingdom}


\begin{abstract}
We compute 4-point correlators in $\mathcal{N} = 4$ $SU(N)$ Super-Yang-Mills 
with both single and double-particle $1/2$-BPS operators in the regime of large ’t 
Hooft coupling and large $N$. In particular we give explicit expressions up to 
$\mathcal{O}(1/N^4)$ for the lightest correlator with two double-particle and 
two single-particle operators built out of the stress-tensor multiplet. 
Our derivation makes use of  the general holographic prescription applied to 
an asymptotically AdS$_5\times$S$^5$ geometry that describes a coherent 
superposition of  multi-graviton operators.
The final result can be written in terms of a natural generalisation of the standard 
D-functions and takes a compact form in Mellin space. The correlator we compute 
here is the simplest of a more general class of correlators where two inserted 
operators are multi-particles. These can be derived with the same approach, 
suggesting that the structure found here is general.
\end{abstract}

\maketitle

\section{Introducing the problem}
\label{sec:intro}

In this letter we study a new class of 4-point correlators in $\mathcal{N}=4$ SYM 
at strong coupling and large values of the central charge. In this regime the theory 
has a dual description in terms of semiclassical type IIB supergravity on AdS$_5\times$S$^5$.  
We focus on 1/2-BPS local operators that can be identified very precisely on both sides 
of the duality. On the SYM side, they belong to the representations of the R-symmetry 
group $SU(4)_R$ with Dynkin labels $[0,p,0]$. In the free theory we can construct all of 
them explicitly by introducing a basis of single and multi-trace operators. Single-traces $T_p$ are defined by
\begin{equation}\label{eq:Top}
T_p(x,y)=y_{I_1}\ldots y_{I_p} \mathrm{Tr}\left( \phi^{I_1}(x) \ldots \phi^{I_p}(x) \right)
\end{equation}
where $\phi^{I}$ are the six elementary real scalars of the ${\cal N}=4$ multiplet 
and $\vec{y}\in \mathbb{C}^6$ is null, $\vec{y}\cdot\vec{y}=0$. Multi-trace operators 
are products of the type $T_{p_1}(x,y)\ldots T_{p_n}(x,y)$ where the single-trace constituents 
have the {\it same} R-symmetry $y$, and thus supersymmetry ensures that 
there are no short-distance singularities when multiplying them.
On the gravity side, the spectrum of quadratic fluctuations of type IIB supergravity 
around AdS$_5 \times$S$^5$~\cite{Kim:1985ez} identifies a preferred class 
of 1/2-BPS {\it single-particle} operators for given rep $[0,p,0]$  that we will 
denote with ${\mathcal O}_p$. In the $SU(N)$ theory, the lightest single-particle is 
a single-trace, ${\cal O}_2=T_2$, of dimension $\Delta=2$. Similarly, the lightest 
double-particle is ${\cal O}_2^2=T_2^2$ of dimension $\Delta=4$. 
The single-particle ${\cal O}_4$ on the other hand is an admixture 
of $T_4$ and $T_2^2$ that is orthogonal to ${\cal O}_2^2$~\cite{Arutyunov:1999en,DHoker:1999jke}, and
more generally, the ${\cal O}_{p\geq 4}$ are uniquely defined 
to be orthogonal to all multi-trace operators \cite{Aprile:2020uxk}. 
Four-point single-particle correlators have been thoroughly studied 
by using Witten diagrams~\cite{Witten:1998qj}
since the early days of the AdS/CFT duality
 see e.g.~\cite{Freedman:1998tz,Freedman:1998bj,DHoker:1999kzh,Arutyunov:2000py,Arutyunov:2002fh,Uruchurtu:2008kp,Arutyunov:2018neq,Arutyunov:2018tvn}.
More recently, the combination of CFT techniques, Ward Identities \cite{Eden:2000bk,Dolan:2001tt}, 
and Operator Product Expansion \cite{Arutyunov:2000ku,Dolan:2006ec}, 
led the authors of \cite{Rastelli:2017udc} to bootstrap a 
single formula for all Kaluza-Klein single-particle correlators, 
whose simplicity was later explained in~\cite{Caron-Huot:2018kta}
as a consequence of an emergent 10d conformal symmetry.
However the spectrum of $\mathcal{N}=4$ SYM contains many 
more 1/2-BPS {\it multi-particle} states, therefore, in order to solve 
the theory at strong coupling, we need to understand with high accuracy 
also 4-point multi-particle correlators. The prototypical example we shall study is
\begin{equation}
  \label{eq:4pdp}
  \!\!\!{\mathcal C}_{(n,q)}\equiv \langle {\cal O}_{\! 2}^n(x_1,y_1) {\cal O}_{\! 2}^n(x_2,y_2) {\cal O}_{\! q}(x_3,y_3) {\cal O}_{\! q}(x_4,y_4) \rangle\,.
\end{equation}
So far very little is known about such correlators:~\cite{Paulos:2012nu,Ma:2022ihn} 
analysed the relevant Witten diagrams in scalar toy models, and~\cite{Bissi:2021hjk} 
provided an analysis of the Ward Identities relevant to correlators with quarter-BPS 
multi-particle operators. Here we address the task of computing multi-particle 
correlators from the gravity side in a full-fledged holographic theory, ${\cal N}=4$ SYM, 
and we report on the first explicit result for Eq.~\eqref{eq:4pdp} with $n=q=2$. 
We do so by following the supergravity approach of~\cite{Ceplak:2021wzz} 
where the multi-particle correlators were discussed in the AdS$_3$/CFT$_2$ context. 
Generalising that approach to AdS$_5$ we are able to provide an explicit expression 
for ${\cal C}_{(2,2)}$, and from this example, isolate the basis of functions needed for 
describing the generic correlators in~\eqref{eq:4pdp} and their structure in Mellin space. 
As evidence that this structure is general, we show that the AdS$_3$ results of~\cite{Ceplak:2021wzz} 
can be greatly simplified by writing them in terms of the basis of functions introduced here.

\section{The supergravity approach to multi-particle correlators}

The key idea is to study the generating function of the correlators \eqref{eq:4pdp} 
for different values of $n$ in the form of the 4-point function $\langle O_H O_H O_L O_L\rangle$, 
where $O_L$ is ${\cal O}_{\! q}$ (or one of its superconformal descendants) 
and  $O_H$ is a one-parameter coherent sum of ${\cal O}_{\! 2}^n$ of the form 
$O_H(\epsilon)=\sum_{n\geq 0} \epsilon^n N^n {\cal O}_2^n$. When 
the sum is peaked around values of $n$ of the order of $N^2$, 
one obtains a Heavy-Heavy-Light-Light (HHLL) correlator 
where the heavy operators, $O_H$, are mapped to a non-trivial 
asymptotically AdS$_5\times$S$^5$ geometry by the holographic dictionary. 
The HHLL correlators can be computed in the dual gravity theory by studying 
the linearised perturbation sourced by the light operator, on top of the geometry 
sourced by the heavy operator. These geometries are the bubbling solutions 
constructed by LLM in the beautiful work~\cite{Lin:2004nb}.
For our purposes it will be sufficient to identify the family of  geometries
dual to $O_H(\epsilon)$ that reduces to a small deformation of AdS$_5\times$S$^5$ 
as $\epsilon\to 0$: in this limit one recovers the gravitational perturbation 
found originally in \cite{Kim:1985ez} that is dual to the light 1/2-BPS single-particle state ${\mathcal O}_2$.

We will demonstrate that the small $\epsilon$ expansion of the HHLL correlator 
encodes explicit expressions for~\eqref{eq:4pdp} with different finite values $n$.

The approach outlined above is part of a general programme that  was successfully 
implemented in the AdS$_3$/CFT$_2$ case by using the 2-charge geometries 
discussed in~\cite{Lunin:2001jy,Maldacena:2000dr,Kanitscheider:2007wq} for 
describing the heavy states: HHLL correlators were derived in~\cite{Galliani:2017jlg,Bombini:2017sge,Bombini:2019vnc} 
and their light limit was studied in~\cite{Giusto:2018ovt} providing the first 
examples of tree-level correlators among single-particle states for that holographic duality. 
The results have been subsequently checked and generalised in~\cite{Rastelli:2019gtj,Giusto:2019pxc,Giusto:2020neo}. 
The computation we present here for AdS$_5$ follows closely the logic used in AdS$_3$: 
in practice, one needs to solve the wave equation for the perturbation dual to the 
light operator finding a solution which is regular in the interior of the geometry 
and whose non-normalisable contribution at the AdS boundary is a delta function 
at the position of a single-trace operator (for instance $x_3$). Then the value of this 
solution for $x_4\not=x_3$ is the supergravity contribution to the correlator~\eqref{eq:4pdp}.

Let us summarise the key points of this supergravity derivation while leaving 
the technical details to an upcoming work~\cite{Aprile:2025hlt}. The geometry dual 
to the coherent state $O_H(\epsilon)$ built from multigravitons operators ${\cal O}_2^n$ 
is described in~\cite{Liu:2007xj,Giusto:2024trt}. The check that, in this geometry, ${\cal O}_2^n$ 
has no mixing with single-particle operators with $\Delta=2n$, was performed 
in~\cite{Giusto:2024trt} by using the technique introduced in~\cite{Skenderis:2006uy,Skenderis:2007yb} 
and again follows closely an analogous calculation done in AdS$_3$~\cite{Ganchev:2023sth}. 
The $SO(2)\times SO(4) \in SO(6)$ symmetry of the gravity solution implies that the 
polarisations of the heavy operators in the HHLL correlator are aligned in a plane: 
$\vec{y}_1=\{1,i,0,0,0,0\}=(\vec{y}_2)^\star$. To study the light perturbation it is convenient 
to work with a descendant rather than the chiral primary ${\cal O}_p$, as the dual wave 
equation is simpler. Thus, for ${\cal O}_2$ we will consider the operator 
$L=Q^4 {\cal O}_2$, where $Q^4$ is the product of the four charges acting non-trivially 
on ${\cal O}_2$. The operator $L$ is dual to the lowest Kaluza-Klein mode of the 
axion-dilaton field, which we will call $\Phi$.
The wave equation for $\Phi$ is simply 
the massless Klein-Gordon equation in 5D:
\begin{equation}\label{eq:box5}
\Box_5 \Phi=0\;,
\end{equation}
where $\Box_5$ is calculated on the background metric of~\cite{Liu:2007xj,Giusto:2024trt}. This metric 
reduces to AdS$_5$ in global coordinates at the conformal boundary. We place the 
delta function of the non-normalisable mode at the north pole of the S$^3$ in AdS$_5$ 
and at $t=0$ (with $t$ the global time). Away from this point, the axion-dilaton decays 
as the normalisable mode at the AdS boundary and we can extract its boundary value, 
$\Phi_B$, which depends only on $t$ and the polar angle $\theta$. In particular, the 
combinations $z=e^{i(t+\theta)} = e^{t_e + i\theta}$, $\bar{z}=e^{i(t-\theta)} = e^{t_e - i\theta}$  
are nothing else that the conformal cross-ratios of the 4-point correlator. We use the following conventions
\begin{equation}
  \label{eq:conf-cross}
    \quad U=(1-z)(1-\bar{z})\;,\quad V=z\bar{z}\;.\\ 
\end{equation}

While we don't have an $\epsilon$-exact solution for the wave equation~\eqref{eq:box5},
it is possible to obtain explicit expressions  by expanding both $\Phi$ and 
$\Box_5$ in \eqref{eq:box5} in the small $\epsilon$ regime \footnote{For convenience we slightly changed conventions with 
respect to~\cite{Giusto:2024trt}: $\epsilon^2_{\rm there} = 8 \epsilon^2_{\rm here}  + 4\epsilon^4_{\rm here}/3 + {\mathcal O}(\epsilon^6)$.}. Interestingly it is possible to arrange the result in the form, 
\begin{equation}
  \label{eq:Phiexp}
  \Phi_B = \frac{1}{12}\mathcal{D}^2 \left[ \frac{1}{U^2} + \epsilon^2  \Psi_B^{(1)}  + \frac{1}{2} \epsilon^4 \Psi_B^{(2)} +\ldots \right]\;,
\end{equation}
where the differential operator $\mathcal{D}$ reads
\begin{equation}
  \label{eq:opwi}
\mathcal{D}\equiv U \partial_U^2 +  V \partial_V^2 + (U+V-1) \,\partial_U \partial_V + 2 \,\partial_U + 2 \,\partial_V {}\,.
\end{equation}
The first terms of the $\epsilon$ expansion are
\begin{align}
  \label{eq:Psi02}
  \Psi_B^{(1)} & = Q^{(1)}_{7} \mathcal{P}_1 + 
  				Q^{(2)}_{6} \log V + 
				Q^{(3)}_{6} \log U + 
				Q^{(4)}_{4}\;, \\  \nonumber
 \Psi_B^{(2)}   &=  R^{(1)}_{11} \mathcal{P}_2 + 
 			R^{(2)}_{10} \left[z \partial_z - \bar{z} \partial_{\bar{z}} \right] \mathcal{P}_2+ 
			R^{(3)}_{11} \mathcal{P}_1 \log V\\ 
		     & + R^{(4)}_{10} \log^2 V+R^{(5)}_{10} \log V\log U   \label{eq:Psi04} \\
 		     & + R^{(6)}_{11} \mathcal{P}_1 +R^{(7)}_{10} \log V+R^{(8)}_{10} \log U +R^{(9)}_{8} \,, \nonumber
\end{align}
where ${\mathcal P}_n$ are the $n$-loop ladder integrals~\cite{Usyukina:1992jd,Usyukina:1993ch}
$$
\mathcal{P}_n = \sum_{r=0}^n \frac{(-1)^r (2n-r)!}{n! (n-r)!\, r!} \log^rV \,\left[\mathrm{Li}_{2n-r}(z)-\mathrm{Li}_{2n-r}(\bar z)\right] 
$$
and the coefficient functions $Q^{(a)}_k$ and $R^{(a)}_k$ 
are rational functions of $z$ and $\bar z$ with denominators 
proportional to $(z-\bar z)^k$. We provide their explicit expressions in 
an ancillary file.  As a first consistency check on our results, 
both ~\eqref{eq:Psi02} and~\eqref{eq:Psi04} are  regular at $z=\bar z$.
It is not a coincidence  that the supergravity result is naturally written in 
terms of the differential operator~\eqref{eq:opwi}. This operator is fixed by 
a superconformal Ward identitiy~\cite{Drummond:2006by,Goncalves:2014ffa} 
that relates the operators ${\cal O}_2$ (with $\vec{y}=\{1,i,0,0,0,0\}$) and 
$L$~\footnote{We thank Congkao Wen for directing us to the relevant references on this subject.}. 
Note that by choosing $\Psi_B^{(n)}$ of the form specified above, in particular by not including 
in $Q^{(4)}_{4}$ and $R^{(9)}_{8}$ regular terms for $z=\bar z$, we have fixed the ambiguity 
associated with the kernel of $\mathcal{D}^2$. We will discuss the meaning of this choice, 
together with further checks on our computation, after the supergravity result is 
re-interpreted as a CFT correlator in the next section.

We conclude this section with some comments on the small $\epsilon$ expansion. 
The coherent state $O_H$ is dual to a non-trivial supergravity geometry when the 
sum over $n$ is peaked at values of order $N^2$ in the large $N$ limit. The ratio $n/N^2$ 
is controlled by $\epsilon$: for small $\epsilon$ one has $n \approx N^2 \epsilon^2$. 
Thus, the regime of validity of the supergravity derivation of the correlator $\langle O_H O_H O_L O_L\rangle$ is when we first take $N$ and the 't~Hooft coupling large and then take $\epsilon\ll 1$, so we are dealing with a heavy state built 
with a large number of single-particle constituents, even if this is a small fraction of $N^2$. 
However when dealing with operators that are individually $1/2$-BPS, this result is also 
valid in the regime where $n$ is kept finite and then $N$ is taken to be large~\footnote{We 
believe that this is due to supersymmetry which probably ensures that the interaction 
between the light and the heavy operator is determined by the interaction between 
the individual constituents of the heavy states and the light operator.}, as needed to 
compute the tree-level contributions to the correlators \eqref{eq:4pdp}. This was first 
noticed in the AdS$_3$ context~\cite{Giusto:2018ovt,Giusto:2019pxc} and can be 
checked starting from~\eqref{eq:Psi02} as done below. Another well-studied scaling 
limit is to take the dimension ($\Delta_H$) of the heavy operators to be large at fixed $g_{YM}^2 \Delta_H$~\cite{Paul:2023rka,Brown:2023why,Caetano:2023zwe,Brown:2024yvt}. 
It would be interesting to study whether also this regime is smoothly connected to the gravity results.

\section{The simplest AdS$_5$ correlator with two double-particle operators}

In order to extract from $\Psi_B^{(n)}$ the correlators \eqref{eq:4pdp}, it is useful to 
recall that, according to the partial non-normalisation theorem~\cite{Eden:2000bk}, 
these correlators can be decomposed as the sum of two pieces, the free correlator 
at $g_{YM}=0$, ${\cal G}_{(n,q)}^\mathrm{free}$, and the genuine dynamical function, ${\cal H}_{(n,q)}$:
\begin{equation}\label{eq:partnonren}  
\hat{{\cal C}}_{(n,q)} \equiv \frac{{\cal C}_{(n,q)}}{
\langle {\cal O}^n_2  {\cal O}^n_2 \rangle \langle {\cal O}_q  {\cal O}_q \rangle} 
=  {\cal G}_{(n,q)}^\mathrm{free}+ {\cal I}\,{\cal H}_{(n,q)}\,,
\end{equation}
where
\begin{equation}
  \label{eq:gij}
    g_{ij} = \frac{\vec{y}_i\cdot \vec{y}_j}{\vec{x}_{ij}^{\,2}}\;,\quad 
   \frac{g_{13} g_{24}}{g_{12}g_{34}} = U \,\sigma\;,\quad  
   \frac{g_{14} g_{23}}{g_{12}g_{34}} = \frac{U}{V}\,\tau\,,
\end{equation}
and
\begin{equation}
  \label{eq:calI}
  \begin{array}{rl}
		{\cal I}\, =&V + \sigma\,V(V-1-U)+\tau\,(1-U-V) \\[.2cm] 
 				&\ + \sigma \tau\, U (U-1-V)+\,\sigma^2 \,U V + \tau^2 \,U\,.
 \end{array}
\end{equation}
When $q=2$, $\mathcal{H}_{(n,2)}$ is a function of only $U$ and $V$. 
This observation allows us to reconstruct the full correlator from the 
supergravity input~\eqref{eq:Psi02}, keeping into account the following three facts:

(i) The choice of polarisations implicit in our supergravity computation 
corresponds to $\sigma=1$, $\tau=0$, and thus $\mathcal{I}=V^2$.

(ii) When $Q^{(4)}_{4}$ and $R^{(9)}_{8}$ are chosen not to contain 
terms that are regular for $z=\bar z$, the supergravity results $\Psi_B^{(n)}$ 
have to be identified with the dynamical part of the correlators  ${\cal C}_{(n,2)}$.

(iii) The supergravity computation captures only the {\it tree-level connected} 
part of the correlator~\footnote{With a slight abuse of nomenclature we
include in the disconnected part the contributions that are
disconnected as seen from the point of view of the 6-point correlator
with single-particles insertions, that is related to ${\mathcal C}_{(2,2)}$ via a
double OPE.}, that we denote by a superscript ``$c$". While for $n=1$ 
the disconnected part is only given by a free contribution, this subtlety 
is important for $n=2$, where the disconnected part includes a non-trivial 1-loop term, 
as it was discussed already in~\cite{Ceplak:2021wzz} in the AdS$_3$ 
context (see for instance Fig.~1 of that paper). By definition $\hat{\cal C}^c_{n}$ 
starts at order $1/N^{2n}$ and our results do not capture the subleading orders.

Then we can identify 
\begin{equation}
{\cal H}^c_{(n,2)} = \frac{1}{a^n} \frac{U^2}{V^2} \Psi_B^{(n)}\,, \quad n\geq 1\,,
\end{equation}
where $a=(N^2-1)$. For $n=0$ the correlator ${\cal C}_{(0,2)}$ 
in~\eqref{eq:partnonren} reduces to the 2-point function in the vacuum, corresponding to the first term of the square parenthesis in~\eqref{eq:Phiexp},
thus  ${\cal H}^c_{(0,2)}$ is trivial.

For $n=1$, using the expressions for $Q^{(a)}_{k}$ given in the ancillary file, 
one can check that this gives 
\begin{equation}
a\, {\cal H}^c_{(1,2)} = -4 U^2 \overline{D}_{2422}\,,
\end{equation}
in agreement with \cite{Dolan:2004iy}, see also \cite[(9.12)]{Dolan:2001tt}. 
This check was recently performed by using a slightly different geometry 
in~\cite{Turton:2024afd} (also for $q > 2$), and supports the claim that we 
can exchange the order of limits, as discussed at the end of the previous section.

For $n=2$, one can write a relatively compact expression by going to Mellin space~\cite{Mack:2009mi,Penedones:2010ue,Fitzpatrick:2011ia,Rastelli:2016nze}:
\begin{equation}\label{eq:H22}
	a^2 \mathcal{H}^c_{(2,2)} = \oint \!\frac{ds dt}{(2\pi i)^2} \,U^{s+2} V^t \,\Gamma^2(-s) \Gamma^2(-t) \Gamma^2(-u) \mathcal{M}^c_{(2,2)}\,,
\end{equation}
where $u=-s-t-4$. The Mellin transform $\mathcal{M}^c_{(2,2)}$ can be 
readily computed from $\Psi_B^{(2)}$ using the approach of~\cite{Aprile:2020luw}:
\begin{equation}\label{eq:Mellin}
	\mathcal{M}^c_{(2,2)}= K(t,u) A + J(t,u) B + C\,,
\end{equation}
where 
\begin{align}
	J(t,u) =&\ \psi^{(0)}(-t)-\psi^{(0)}(-u)\;, \\
	K(t,u) =&\ \psi^{(1)}(-t)+\psi^{(1)}(-u)-\big(J(t,u)\big)^2 -\pi^2\,.
\end{align}
and the functions $A$, $B$ and $C$ are quite simple:
$$
\begin{aligned} 
	A&= \frac{32}{(s+1)(s+2)(s+3)}\Big[ (-16 - 3 s + 3 tu) \\ & \quad + \frac{18 tu }{(s+4)} - \frac{18 t^2 u^2}{(s+4)(s+5)}\Big]\,,\\
	B&= \frac{32(t-u)}{(s+1)(s+2)}\! \left[\frac{18(\,(s+5)-tu)}{(s+4)(s+5)}-\frac{1}{(t+1)(u+1)}\right], \\
	C&= \frac{16}{(s+1)}\!\left[\frac{3(s^2+17 s+64 - 12 tu )}{(s+4)(s+5)}-\frac{1}{(t+1)(u+1)}\right] .
\end{aligned}
$$
The functions $K$ and $J$ correspond in position space to uniform 
transcendentality 4~\cite{Allendes:2012mr} and 3, respectively, and 
are not present in the tree-level single-particle correlators, rather they 
first appear in the 1-loop correlator $\langle \mathcal{O}_2 \mathcal{O}_2\mathcal{O}_2\mathcal{O}_2\rangle$ \cite{Alday:2017xua,Aprile:2017bgs,Alday:2017vkk,Alday:2018kkw,Aprile:2019rep}, 
whose structure is similar to \eqref{eq:Mellin}, apart from being symmetric 
under the exchange of all three Mellin variables, as implied by full crossing 
symmetry of $\langle \mathcal{O}_2 \mathcal{O}_2\mathcal{O}_2\mathcal{O}_2\rangle$. 
Instead~\eqref{eq:Mellin} is symmetric only under $t \leftrightarrow u$ corresponding 
to exchange of the two double-particle (or the two single-particle) operators. 

From ${\cal H}^c_{(2,2)}$ and the free part
\begin{equation}
	 {\cal G}_{(2,2)}^\mathrm{free,c}=
	 \frac{8}{a^2} \left[ 2 U \sigma + 2 \frac{U}{V} \tau + U^2 \sigma^2 +6 \frac{U^2}{V} \sigma \tau + \frac{U^2}{V^2} \tau^2\right]\,,
\end{equation}
we can re-construct the full $n=q=2$ correlator~\eqref{eq:4pdp} 
up to order $1/a^2$, simply by relating the connected 
and the full generating functionals for the correlators in the usual way~\footnote{${\mathcal W} 
=\ln {\mathcal Z}$ where ${\cal W}$ is the connected and ${\cal Z}$ is the full generating functional.}. 
We obtain 
\begin{equation}
  \label{eq:Ccompl}
  \hat{{\mathcal C}}_{(2,2)} = \hat{{\mathcal C}}^{\rm c}_{(2,2)} + \frac{2a}{a+2} \hat{\cal C}_{(1,2)}+\ldots\;,
\end{equation}
where the dots stand for terms that will not contribute to $\mathcal{H}_{(2,2)}$ 
and the $a$-dependence of the disconnected contribution follows from the definition \eqref{eq:Top}, which implies:
\begin{align}
  \label{eq:O2nNor}
 \langle {\mathcal O}_2  {\mathcal O}_2 \rangle = 2 a g_{34}^2 \,\,\,, \,\,\,  
 	\langle {\mathcal O}^2_2  {\mathcal O}^2_2 \rangle = 8 a (a+2) g_{12}^{4} \;.
\end{align}
Through $\hat{\cal C}_{(1,2)}$ we see that both the tree level and 1-loop results for 
$\langle \mathcal{O}_2 \mathcal{O}_2\mathcal{O}_2\mathcal{O}_2\rangle$ (that 
for completeness we include in the supplemental material) enter the full correlator 
as a disconnected contribution.
Then, 
\begin{align}
\label{eq:Hdynafdisco}
  \left[\mathcal{H}_{(2,2)}\right]_{\frac{1}{a}}\ =&\ 2 \left[{\cal H}_{(1,2)}\right]_{\frac{1}{a}} \;,\\[.2cm]
  \label{eq:Hdynaf}
  \left[ \mathcal{H}_{(2,2)} \right]_{\frac{1}{a^2}}=&\  \mathcal{H}^c_{(2,2)} -4 \left[\mathcal{H}_{(1,2)}\right]_{\frac{1}{a}} + 2 \left[\mathcal{H}_{(1,2)}\right]_{\frac{1}{a^2}} \;,
\end{align}
where the square parenthesis select a particular term in the $1/a$ expansion.

{\bf Checks.} We performed several consistency checks on our results, 
by looking at the 3-point couplings of protected operators exchanged in different 
OPE limits. In the easiest instances, the protected exchanges are reproduced 
by just the free part of the correlator. For example, in the $x_1\to x_2$ OPE 
channel, where we look at the common OPE between ${\cal O}_2\times {\cal O}_2$ and ${\cal O}_2^2\times {\cal O}_2^2$, 
the only twist 2 operator exchanged in the $[0,2,0]$ rep is $\mathcal{O}_2$ 
at spin 0, plus the R-current and the stress-energy tensor, which sit in the same
graviton multiplet, at higher spins. One can check that the 3-point couplings of these 
operators agree with those extracted from  ${\cal G}_{(2,2)}^\mathrm{free,c}$, 
and this shows that the dynamical part, ${\cal H}^c_{(2,2)}$, has been correctly 
identified. In the $[0,0,0]$ rep, on the other hand, the twist 2, spin $l$ operators must 
be absent since they are stringy states \cite{Arutyunov:2000ku,Dolan:2006ec}, and indeed one 
verifies that the free and dynamical parts cancel each other. 
Instead, the study of the OPE channel $x_2\rightarrow x_3$, where we look at the common 
OPE between ${\cal O}_2\times {\cal O}^2_2$ and ${\cal O}_2\times {\cal O}_2^2$, is 
much more non-trivial because there are protected double-trace operators 
of twist $4$ in the $[0,2,0]$ rep 
whose 3-point couplings are not readily available. 
However, in the $[0,2,0]$ rep 
these twist $4$ protected  operators   are the only ones
exchanged in the ${\cal O}_2\times{\cal O}_2$ OPE, 
and therefore, as explained in \cite{Doobary:2015gia}, their contribution to 
${\cal C}_{(2,2)}$ is computed by requiring that the $2\times2$ Gram 
matrix of 4-pt functions $\{ \langle {\cal O}_2 {\cal O}_2|,\langle {\cal O}_2 {\cal O}^2_2|\}\otimes\{ | {\cal O}_2 {\cal O}_2\rangle,| {\cal O}_2 {\cal O}^2_2\rangle\}$ 
has vanishing determinant 
once projected onto the corresponding rep.
This computation only needs free theory, is exact in $a$, and non 
trivial in spin.
Then, the same argument as in \cite{Aprile:2019rep} implies a prediction 
for ${\cal H}_{(2,2)}$ at twist $4$ in the orientation $\langle {\cal O}_2{\cal O}_2^2 {\cal O}_2 {\cal O}_2^2\rangle$. See also \cite{Bissi:2024tqf}.
The order $1/a$ prediction is actually a non-standard check for the 
tree level disconnected correlator \eqref{eq:Hdynafdisco}, which we 
will discuss in detail in \cite{Aprile:2025hlt}. 
The $1/a^2$ prediction  is
\begin{equation}
  \label{eq:blockdec}
  \bigl[\mathcal{H}_{(2,2)}\bigr]_{\frac{1}{a^2}} = 64 \sum_{l\in 2 \mathbb{N}}  (l+2)(l+5)\frac{ (l+2)!^2}{(2l+6)!}\, {\tt Blc}_{\tau=4,l}+\ldots
\end{equation}
where we have not included higher than twist four contributions in the block 
expansion on the rhs, and the blocks ${\tt Blc}_{\tau,l}$ are given in \cite[(A.33)]{Aprile:2020mus}. 
Quite remarkably, we verified that~\eqref{eq:Hdynaf} 
is consistent with this CFT prediction for all spins, and moreover 
we have found that this check could not possibly work without ${\cal H}_{(2,2)}^c$.

\section{Outlook} \label{sec:outlook}

We obtained the first explicit result for a holographic 4-point correlator in 
${\mathcal N}=4$ SYM involving double-particle operators. We found that the 
connected correlator is a function of trascendentality up to four, where the 
double-box ladder integral ${\mathcal P}_2$ features as the main character. 
We believe this structure applies beyond the AdS$_5$ case. As a first evidence 
of this, one can  compare with the AdS$_3$/CFT$_2$ results.
In this case the lowest CPO has dimension $\Delta=1$, so the first connected 
correlator has $n=1$, $q=1$. Then, multi-particle correlators of the form \eqref{eq:4pdp} 
have been computed in \cite{Ceplak:2021wzz} for $n=2,3,4$. From the results 
in \cite{Ceplak:2021wzz} we found that all these correlators can be written just 
in terms of ${\cal P}_n$ and their derivatives, similarly to what happens in 
\eqref{eq:Psi02}-\eqref{eq:Psi04}. Here one can try to take a step further 
and ask if these correlators  admit a compact re-writing analogous 
to the case of $n=1$ with a single D-function.  A natural generalisation 
of the latter for the class of correlators in \eqref{eq:4pdp} would be 
\begin{equation}
D^{(n)}(x_i)\equiv \frac{1}{(x^2_{12})^{n} \,x^2_{34}} \frac{|1-z|^2}{z-\bar{z}}\,\mathcal{P}_n(z,\bar{z})\,.
\end{equation}
The $x_{ij}^2$-derivatives of $D^{(1)}$ are indeed the usual D-functions
describing the 4-point tree level correlators. Remarkably we find that 
$x_{ij}^2$-derivatives of $D^{(n)}$ play the same role for correlators of 
the type~\eqref{eq:4pdp} with $n>1$. For instance, Eq.~(3.8c) of~\cite{Ceplak:2021wzz} 
can be written as~\footnote{In AdS$_3$ $N$ has a different meaning than 
in ${\mathcal N}=4$ SYM and is proportional to the central charge $N \sim a$.}
\begin{equation}\label{eq:ads3two}
:\! C_2\! : = -\frac{2 \,|z|^2}{N^2}\,\left[\frac{\partial D^{(2)}}{\partial x^2_{34}} - (z+\bar{z})\,\frac{\partial^2 D^{(2)}}{\partial (x^2_{34})^2}  - \frac{\partial D^{(1)}}{\partial x^2_{34}}\right]\,.
\end{equation}
For the multi-particle correlators with $n=3,4$ in \cite{Ceplak:2021wzz}
we found analogous compact expressions in terms of $D^{(n)}$ with $n\le 4$! 
Going back to AdS$_5$ we  checked that one can write ${\cal H}^c_{(2,2)}$ 
using derivatives of $D^{(2)}$ and $D^{(1)}$. The existence of such rewritings 
hints at the existence of a differential representation 
similar to the one observed for certain single particle correlators 
\cite{Huang:2024dck,Huang:2024rxr}.
It also hints at the existence of recursion relations both 
in $n$ and in $q$. The latter may be related to the possibility of 
extending the 10D conformal symmetry of~\cite{Caron-Huot:2018kta}
to the class of multiparticle correlators discussed in this letter.
It would also be interesting to study stringy corrections, and 
make a precise connection with integrated HHLL correlators~\cite{Paul:2023rka,Brown:2023why}.
 Another direction where our results can be useful is the study of higher points 
 holographic correlators, since these reduce to correlators with multi-particle 
 operators in the appropriate OPE limit. For example, the 5-point correlator of 
 ${\cal O}_2$ can be written in terms of standard D-functions~\cite{Goncalves:2019znr,Goncalves:2023oyx}, 
 but our results shows that this is not the case in general, starting from 6-point correlators. 
In this sense, our method could be used together with other approaches, 
such as the one in~\cite{Alday:2023kfm,Cao:2023cwa}, to derive explicit expressions for 
tree-level $n$-point correlators. On the CFT side, the correlators presented 
here contain new CFT data for long triple-particle operators in ${\mathcal N}=4$ 
SYM at strong coupling, and one could use them as an input to extend the 
unmixing programme started in~\cite{Aprile:2017xsp}.
 Finally, it would be very interesting to apply the supergravity approach used 
 here to other AdS/CFT dualities in diverse dimensions that admit a regime well 
 approximated by supergravity, such as ABJM~\cite{Aharony:2008ug}.

{\bf Acknowledgements.}
It is a pleasure to acknowledge fruitful scientific exchange with 
F.~Galvagno, A.~Georgoudis, H.~Paul, A.~Tyukov, D.~Turton, C.~Wen.
We would like to thank A.~Bissi e G~Fardelli for helpful comments on the draft, 
 insightful discussions and correspondence.
F.~Aprile is supported by the Ramon y Cajal program through the 
fellowship RYC2021-031627-I funded by MCIN/AEI/10.13039/501100011033 
and by the European Union NextGenerationEU/PRTR. 
R.~Russo. is partially supported by the UK EPSRC grant ``CFT and 
Gravity: Heavy States and Black Holes'' EP/W019663/1 and 
the STFC grants ``Amplitudes, Strings and Duality'', 
grant numbers ST/T000686/1 and ST/X00063X/1. 
No new data were generated or analysed during this study.

\newpage

\begin{widetext}
\begin{center}
  {\bf SUPPLEMENTAL MATERIAL}
\end{center}
We emphasised that the gravity computation only captures the tree-level connected correlator between two multi-trace and two single-trace operators. We give here for completeness the complete definition of the connected correlator with two double-trace operators, thus making explicit the dots in \eqref{eq:Ccompl}:
\begin{equation}
\begin{aligned}
&{\mathcal C}^{c}_{(2,2)}\equiv  \langle {\cal O}_{\! 2}^2(1) {\cal O}_{\! 2}^2(2) {\cal O}_{\! 2}(3) {\cal O}_{\! 2}(4)\rangle^{c}=\langle {\cal O}_{\! 2}^2(1) {\cal O}_{\! 2}^2(2) {\cal O}_{\! 2}(3) {\cal O}_{\! 2}(4)\rangle - 4 \langle {\cal O}_{\! 2}(1) {\cal O}_{\! 2}(2) \rangle \langle {\cal O}_{\! 2}(1) {\cal O}_{\! 2}(2) {\cal O}_{\! 2}(3) {\cal O}_{\! 2}(4)\rangle \\
&-2\Bigl(\langle {\cal O}_{\! 2}(1) {\cal O}_{\! 2}(3) \rangle \langle {\cal O}_{\! 2}(1) {\cal O}_{\! 2}^2(2) {\cal O}_{\! 2}(4)\rangle + \langle{\cal O}_{\! 2}(1) {\cal O}_{\! 2}(4) \rangle \langle {\cal O}_{\! 2}(1) {\cal O}_{\! 2}^2(2) {\cal O}_{\! 2}(3)\rangle + \langle{\cal O}_{\! 2}(2) {\cal O}_{\! 2}(3) \rangle \langle {\cal O}_{\! 2}^2(1) {\cal O}_{\! 2}(2) {\cal O}_{\! 2}(4)\rangle\\
&  +\langle {\cal O}_{\! 2}(1) {\cal O}_{\! 2}(4) \rangle \langle {\cal O}_{\! 2}^2(1) {\cal O}_{\! 2}(2) {\cal O}_{\! 2}(3)\rangle   \Bigr)-4 \langle {\cal O}_{\! 2}(1) {\cal O}_{\! 2}(2) {\cal O}_{\! 2}(3) \rangle \langle {\cal O}_{\! 2}(1) {\cal O}_{\! 2}(2) {\cal O}_{\! 2}(4) \rangle-\langle {\cal O}_{\! 2}^2(1) {\cal O}_{\! 2}^2(2)  \rangle \langle  {\cal O}_{\! 2}(3) {\cal O}_{\! 2}(4)\rangle \\
&+ 4 (\langle {\cal O}_{\! 2}(1) {\cal O}_{\! 2}(2)  \rangle)^2 \langle  {\cal O}_{\! 2}(3) {\cal O}_{\! 2}(4)\rangle+8 \langle {\cal O}_{\! 2}(1) {\cal O}_{\! 2}(2) \rangle \left(\langle {\cal O}_{\! 2}(1) {\cal O}_{\! 2}(3)\rangle  \langle {\cal O}_{\! 2}(2) {\cal O}_{\! 2}(4)\rangle+\langle {\cal O}_{\! 2}(1) {\cal O}_{\! 2}(4)\rangle  \langle {\cal O}_{\! 2}(2) {\cal O}_{\! 2}(3)\rangle\right) \,.
\end{aligned}
\end{equation}
The Mellin transform of the normalised connected correlator, encoded in its dynamical part, $\mathcal{H}^c_{(2,2)}$, has been given in \eqref{eq:Mellin}. The Mellin of $\mathcal{H}_{(1,2)}$, that captures the contribution of the last term in the first line of the equation above, has a fully crossing-symmetric form given by
\begin{equation}\label{eq:Mellin1loop}
\mathcal{M}_{(1,2)}=\frac{4}{a} D_\mathrm{d} +\frac{16}{a^2} \left[\big( K(t,u) A_\mathrm{d} +\mathrm{sym.} \big)+ \big(\psi^{(0)}(-u)B_\mathrm{d} + \mathrm{sym.}\big)  + \big(C_\mathrm{d} +\mathrm{sym.}\big) \right]\,,
\end{equation}
with
\begin{align}
D_\mathrm{d}(s,t) &= \frac{1}{(s+1)(t+1)(u+1)}
\end{align}
and
\begin{align}
A_\mathrm{d} &=\frac{1}{20}\Bigg[ \frac{ t^2 u^2 \big( -15- \frac{12}{s+4}+\frac{72}{s+5} \big) }{(s+1)_3} + \frac{tu \big( 25 -\frac{6}{s+3}-\frac{60}{s+4}\big) }{(s+1)_2} -\frac{4\big(3-\frac{2}{s+2}-\frac{1}{s+3}\big)}{(s+1)_1}\Bigg] \\
B_\mathrm{d} &=\frac{1}{20}\Bigg[ (t-u)\Bigg[ \frac{ tu \big( 15+\frac{12}{s+4}-\frac{72}{s+5} \big) }{(s+1)(s+2) } +10-\frac{2}{s+1}-\frac{8}{s+2}-\frac{10}{s+4}  \Bigg]\Bigg]+ (t\leftrightarrow s)\\
C_\mathrm{d} &=\frac{1}{20}\Bigg[ \frac{tu\big( -15-\frac{12}{s+4}+\frac{72}{s+5}\big)}{(s+1)}+{\frac{2}{(s+1)}-\frac{20}{(s+4)}+\frac{6}{(s+5)}}{}\Bigg] +\frac{1}{4}
\end{align}
The functions $A_d$ and $B_d$ are fixed by the double logarithmic discontinuity and come with poles at and below the unitarity bound, eg.~$s=\{-1,\ldots, -5\}$ etc\ldots. Regarding $C_d$, the $s$ and $t$ dependence is fixed by requiring that the total residue at these poles vanishes, and by consistency with the flat space limit of Penedones \cite{Penedones:2010ue}, see~\cite{Aprile:2020luw}. The only ambiguity left is thus a constant, and in particular the $+\frac{1}{4}$ in $C_d$ is fixed by considering the input from localization and the integrated correlator \cite{Chester:2019pvm}.
Note that $B(s,t)+ B(u,s)+B(t,u)=0$ making possible to rewrite the function in terms of $(\psi(-s)-\psi(-t))$, $(\psi(-s)-\psi(-u))$, $(\psi(-t)-\psi(-u))$, as for ${\cal M}_{(2,2)}$.\\

Ref.~\cite{Ceplak:2021wzz} derives AdS$_3$ correlators of the form \eqref{eq:4pdp} with $n$ up to four (see Eqs. (A.12d)-(A.12e)). We found that all these correlators can be rewritten in a form analogous to \eqref{eq:ads3two}: 
\begin{equation}
\begin{aligned}
:\!C_3\!: &= -\frac{6 \,|z|^2}{N^3}\,\Bigl[\frac{ \partial D^{(3)}}{\partial x_{34}^2} -3 (z+{\bar z})\,\frac{\partial^2 D^{(3)}}{\partial(x_{34}^2)^2} + (z^2+{\bar z}^2+4|z|^2)\,\frac{\partial^3 D^{(3)}}{\partial(x_{34}^2)^3} -\frac{\partial D^{(2)}}{\partial x_{34}^2} + (z+{\bar z})\,\frac{\partial^2 D^{(2)}}{\partial(x_{34}^2)^2} \Bigr]\,,\\
:\!C_4\!: &= -\frac{24 \,|z|^2}{N^4}\,\Bigl[\frac{ \partial D^{(4)}}{\partial x_{34}^2} -7 (z+{\bar z})\,\frac{\partial^2 D^{(4)}}{\partial(x_{34}^2)^2} +  (6(z^2+{\bar z}^2)+28|z|^2)\,\frac{\partial^3 D^{(4)}}{\partial(x_{34}^2)^3}  \\
&- (z^3+{\bar z}^3+11 |z|^2(z+{\bar z}))\,\frac{\partial^4 D^{(4)}}{\partial(x_{34}^2)^4} + 3 (z+{\bar z})\,\frac{\partial^2 D^{(3)}}{\partial(x_{34}^2)^2} - (z^2+{\bar z}^2+4|z|^2)\,\frac{\partial^3 D^{(3)}}{\partial(x_{34}^2)^3} \Bigr]\,.
\end{aligned}
\end{equation}
\end{widetext}


\begin{thebibliography}{72}
\expandafter\ifx\csname natexlab\endcsname\relax\def\natexlab#1{#1}\fi
\expandafter\ifx\csname bibnamefont\endcsname\relax
  \def\bibnamefont#1{#1}\fi
\expandafter\ifx\csname bibfnamefont\endcsname\relax
  \def\bibfnamefont#1{#1}\fi
\expandafter\ifx\csname citenamefont\endcsname\relax
  \def\citenamefont#1{#1}\fi
\expandafter\ifx\csname url\endcsname\relax
  \def\url#1{\texttt{#1}}\fi
\expandafter\ifx\csname urlprefix\endcsname\relax\def\urlprefix{URL }\fi
\providecommand{\bibinfo}[2]{#2}
\providecommand{\eprint}[2][]{\url{#2}}

\bibitem[{\citenamefont{Kim et~al.}(1985)\citenamefont{Kim, Romans, and van Nieuwenhuizen}}]{Kim:1985ez}
\bibinfo{author}{\bibfnamefont{H.~J.} \bibnamefont{Kim}}, \bibinfo{author}{\bibfnamefont{L.~J.} \bibnamefont{Romans}}, \bibnamefont{and} \bibinfo{author}{\bibfnamefont{P.}~\bibnamefont{van Nieuwenhuizen}}, \bibinfo{journal}{Phys. Rev. D} \textbf{\bibinfo{volume}{32}}, \bibinfo{pages}{389} (\bibinfo{year}{1985}).

\bibitem[{\citenamefont{Arutyunov and Frolov}(2000{\natexlab{a}})}]{Arutyunov:1999en}
\bibinfo{author}{\bibfnamefont{G.}~\bibnamefont{Arutyunov}} \bibnamefont{and} \bibinfo{author}{\bibfnamefont{S.}~\bibnamefont{Frolov}}, \bibinfo{journal}{Phys. Rev. D} \textbf{\bibinfo{volume}{61}}, \bibinfo{pages}{064009} (\bibinfo{year}{2000}{\natexlab{a}}), \eprint{hep-th/9907085}.

\bibitem[{\citenamefont{D'Hoker et~al.}(1999{\natexlab{a}})\citenamefont{D'Hoker, Freedman, Mathur, Matusis, and Rastelli}}]{DHoker:1999jke}
\bibinfo{author}{\bibfnamefont{E.}~\bibnamefont{D'Hoker}}, \bibinfo{author}{\bibfnamefont{D.~Z.} \bibnamefont{Freedman}}, \bibinfo{author}{\bibfnamefont{S.~D.} \bibnamefont{Mathur}}, \bibinfo{author}{\bibfnamefont{A.}~\bibnamefont{Matusis}}, \bibnamefont{and} \bibinfo{author}{\bibfnamefont{L.}~\bibnamefont{Rastelli}}, pp. \bibinfo{pages}{332--360} (\bibinfo{year}{1999}{\natexlab{a}}), \eprint{hep-th/9908160}.

\bibitem[{\citenamefont{Aprile et~al.}(2020{\natexlab{a}})\citenamefont{Aprile, Drummond, Heslop, Paul, Sanfilippo, Santagata, and Stewart}}]{Aprile:2020uxk}
\bibinfo{author}{\bibfnamefont{F.}~\bibnamefont{Aprile}}, \bibinfo{author}{\bibfnamefont{J.~M.} \bibnamefont{Drummond}}, \bibinfo{author}{\bibfnamefont{P.}~\bibnamefont{Heslop}}, \bibinfo{author}{\bibfnamefont{H.}~\bibnamefont{Paul}}, \bibinfo{author}{\bibfnamefont{F.}~\bibnamefont{Sanfilippo}}, \bibinfo{author}{\bibfnamefont{M.}~\bibnamefont{Santagata}}, \bibnamefont{and} \bibinfo{author}{\bibfnamefont{A.}~\bibnamefont{Stewart}}, \bibinfo{journal}{JHEP} \textbf{\bibinfo{volume}{11}}, \bibinfo{pages}{072} (\bibinfo{year}{2020}{\natexlab{a}}), \eprint{2007.09395}.

\bibitem[{\citenamefont{Witten}(1998)}]{Witten:1998qj}
\bibinfo{author}{\bibfnamefont{E.}~\bibnamefont{Witten}}, \bibinfo{journal}{Adv. Theor. Math. Phys.} \textbf{\bibinfo{volume}{2}}, \bibinfo{pages}{253} (\bibinfo{year}{1998}), \eprint{hep-th/9802150}.

\bibitem[{\citenamefont{Freedman et~al.}(1999{\natexlab{a}})\citenamefont{Freedman, Mathur, Matusis, and Rastelli}}]{Freedman:1998tz}
\bibinfo{author}{\bibfnamefont{D.~Z.} \bibnamefont{Freedman}}, \bibinfo{author}{\bibfnamefont{S.~D.} \bibnamefont{Mathur}}, \bibinfo{author}{\bibfnamefont{A.}~\bibnamefont{Matusis}}, \bibnamefont{and} \bibinfo{author}{\bibfnamefont{L.}~\bibnamefont{Rastelli}}, \bibinfo{journal}{Nucl. Phys. B} \textbf{\bibinfo{volume}{546}}, \bibinfo{pages}{96} (\bibinfo{year}{1999}{\natexlab{a}}), \eprint{hep-th/9804058}.

\bibitem[{\citenamefont{Freedman et~al.}(1999{\natexlab{b}})\citenamefont{Freedman, Mathur, Matusis, and Rastelli}}]{Freedman:1998bj}
\bibinfo{author}{\bibfnamefont{D.~Z.} \bibnamefont{Freedman}}, \bibinfo{author}{\bibfnamefont{S.~D.} \bibnamefont{Mathur}}, \bibinfo{author}{\bibfnamefont{A.}~\bibnamefont{Matusis}}, \bibnamefont{and} \bibinfo{author}{\bibfnamefont{L.}~\bibnamefont{Rastelli}}, \bibinfo{journal}{Phys. Lett. B} \textbf{\bibinfo{volume}{452}}, \bibinfo{pages}{61} (\bibinfo{year}{1999}{\natexlab{b}}), \eprint{hep-th/9808006}.

\bibitem[{\citenamefont{D'Hoker et~al.}(1999{\natexlab{b}})\citenamefont{D'Hoker, Freedman, Mathur, Matusis, and Rastelli}}]{DHoker:1999kzh}
\bibinfo{author}{\bibfnamefont{E.}~\bibnamefont{D'Hoker}}, \bibinfo{author}{\bibfnamefont{D.~Z.} \bibnamefont{Freedman}}, \bibinfo{author}{\bibfnamefont{S.~D.} \bibnamefont{Mathur}}, \bibinfo{author}{\bibfnamefont{A.}~\bibnamefont{Matusis}}, \bibnamefont{and} \bibinfo{author}{\bibfnamefont{L.}~\bibnamefont{Rastelli}}, \bibinfo{journal}{Nucl. Phys. B} \textbf{\bibinfo{volume}{562}}, \bibinfo{pages}{353} (\bibinfo{year}{1999}{\natexlab{b}}), \eprint{hep-th/9903196}.

\bibitem[{\citenamefont{Arutyunov and Frolov}(2000{\natexlab{b}})}]{Arutyunov:2000py}
\bibinfo{author}{\bibfnamefont{G.}~\bibnamefont{Arutyunov}} \bibnamefont{and} \bibinfo{author}{\bibfnamefont{S.}~\bibnamefont{Frolov}}, \bibinfo{journal}{Phys. Rev. D} \textbf{\bibinfo{volume}{62}}, \bibinfo{pages}{064016} (\bibinfo{year}{2000}{\natexlab{b}}), \eprint{hep-th/0002170}.

\bibitem[{\citenamefont{Arutyunov et~al.}(2003)\citenamefont{Arutyunov, Dolan, Osborn, and Sokatchev}}]{Arutyunov:2002fh}
\bibinfo{author}{\bibfnamefont{G.}~\bibnamefont{Arutyunov}}, \bibinfo{author}{\bibfnamefont{F.~A.} \bibnamefont{Dolan}}, \bibinfo{author}{\bibfnamefont{H.}~\bibnamefont{Osborn}}, \bibnamefont{and} \bibinfo{author}{\bibfnamefont{E.}~\bibnamefont{Sokatchev}}, \bibinfo{journal}{Nucl. Phys. B} \textbf{\bibinfo{volume}{665}}, \bibinfo{pages}{273} (\bibinfo{year}{2003}), \eprint{hep-th/0212116}.

\bibitem[{\citenamefont{Uruchurtu}(2009)}]{Uruchurtu:2008kp}
\bibinfo{author}{\bibfnamefont{L.~I.} \bibnamefont{Uruchurtu}}, \bibinfo{journal}{JHEP} \textbf{\bibinfo{volume}{03}}, \bibinfo{pages}{133} (\bibinfo{year}{2009}), \eprint{0811.2320}.

\bibitem[{\citenamefont{Arutyunov et~al.}(2018{\natexlab{a}})\citenamefont{Arutyunov, Klabbers, and Savin}}]{Arutyunov:2018neq}
\bibinfo{author}{\bibfnamefont{G.}~\bibnamefont{Arutyunov}}, \bibinfo{author}{\bibfnamefont{R.}~\bibnamefont{Klabbers}}, \bibnamefont{and} \bibinfo{author}{\bibfnamefont{S.}~\bibnamefont{Savin}}, \bibinfo{journal}{JHEP} \textbf{\bibinfo{volume}{09}}, \bibinfo{pages}{023} (\bibinfo{year}{2018}{\natexlab{a}}), \eprint{1806.09200}.

\bibitem[{\citenamefont{Arutyunov et~al.}(2018{\natexlab{b}})\citenamefont{Arutyunov, Klabbers, and Savin}}]{Arutyunov:2018tvn}
\bibinfo{author}{\bibfnamefont{G.}~\bibnamefont{Arutyunov}}, \bibinfo{author}{\bibfnamefont{R.}~\bibnamefont{Klabbers}}, \bibnamefont{and} \bibinfo{author}{\bibfnamefont{S.}~\bibnamefont{Savin}}, \bibinfo{journal}{JHEP} \textbf{\bibinfo{volume}{09}}, \bibinfo{pages}{118} (\bibinfo{year}{2018}{\natexlab{b}}), \eprint{1808.06788}.

\bibitem[{\citenamefont{Eden et~al.}(2001)\citenamefont{Eden, Petkou, Schubert, and Sokatchev}}]{Eden:2000bk}
\bibinfo{author}{\bibfnamefont{B.}~\bibnamefont{Eden}}, \bibinfo{author}{\bibfnamefont{A.~C.} \bibnamefont{Petkou}}, \bibinfo{author}{\bibfnamefont{C.}~\bibnamefont{Schubert}}, \bibnamefont{and} \bibinfo{author}{\bibfnamefont{E.}~\bibnamefont{Sokatchev}}, \bibinfo{journal}{Nucl. Phys. B} \textbf{\bibinfo{volume}{607}}, \bibinfo{pages}{191} (\bibinfo{year}{2001}), \eprint{hep-th/0009106}.

\bibitem[{\citenamefont{Dolan and Osborn}(2002)}]{Dolan:2001tt}
\bibinfo{author}{\bibfnamefont{F.~A.} \bibnamefont{Dolan}} \bibnamefont{and} \bibinfo{author}{\bibfnamefont{H.}~\bibnamefont{Osborn}}, \bibinfo{journal}{Nucl. Phys. B} \textbf{\bibinfo{volume}{629}}, \bibinfo{pages}{3} (\bibinfo{year}{2002}), \eprint{hep-th/0112251}.

\bibitem[{\citenamefont{Arutyunov et~al.}(2000)\citenamefont{Arutyunov, Frolov, and Petkou}}]{Arutyunov:2000ku}
\bibinfo{author}{\bibfnamefont{G.}~\bibnamefont{Arutyunov}}, \bibinfo{author}{\bibfnamefont{S.}~\bibnamefont{Frolov}}, \bibnamefont{and} \bibinfo{author}{\bibfnamefont{A.~C.} \bibnamefont{Petkou}}, \bibinfo{journal}{Nucl. Phys. B} \textbf{\bibinfo{volume}{586}}, \bibinfo{pages}{547} (\bibinfo{year}{2000}), \bibinfo{note}{[Erratum: Nucl.Phys.B 609, 539--539 (2001)]}, \eprint{hep-th/0005182}.

\bibitem[{\citenamefont{Dolan et~al.}(2006)\citenamefont{Dolan, Nirschl, and Osborn}}]{Dolan:2006ec}
\bibinfo{author}{\bibfnamefont{F.~A.} \bibnamefont{Dolan}}, \bibinfo{author}{\bibfnamefont{M.}~\bibnamefont{Nirschl}}, \bibnamefont{and} \bibinfo{author}{\bibfnamefont{H.}~\bibnamefont{Osborn}}, \bibinfo{journal}{Nucl. Phys. B} \textbf{\bibinfo{volume}{749}}, \bibinfo{pages}{109} (\bibinfo{year}{2006}), \eprint{hep-th/0601148}.

\bibitem[{\citenamefont{Rastelli and Zhou}(2018)}]{Rastelli:2017udc}
\bibinfo{author}{\bibfnamefont{L.}~\bibnamefont{Rastelli}} \bibnamefont{and} \bibinfo{author}{\bibfnamefont{X.}~\bibnamefont{Zhou}}, \bibinfo{journal}{JHEP} \textbf{\bibinfo{volume}{04}}, \bibinfo{pages}{014} (\bibinfo{year}{2018}), \eprint{1710.05923}.

\bibitem[{\citenamefont{Caron-Huot and Trinh}(2019)}]{Caron-Huot:2018kta}
\bibinfo{author}{\bibfnamefont{S.}~\bibnamefont{Caron-Huot}} \bibnamefont{and} \bibinfo{author}{\bibfnamefont{A.-K.} \bibnamefont{Trinh}}, \bibinfo{journal}{JHEP} \textbf{\bibinfo{volume}{01}}, \bibinfo{pages}{196} (\bibinfo{year}{2019}), \eprint{1809.09173}.

\bibitem[{\citenamefont{Paulos et~al.}(2012)\citenamefont{Paulos, Spradlin, and Volovich}}]{Paulos:2012nu}
\bibinfo{author}{\bibfnamefont{M.~F.} \bibnamefont{Paulos}}, \bibinfo{author}{\bibfnamefont{M.}~\bibnamefont{Spradlin}}, \bibnamefont{and} \bibinfo{author}{\bibfnamefont{A.}~\bibnamefont{Volovich}}, \bibinfo{journal}{JHEP} \textbf{\bibinfo{volume}{08}}, \bibinfo{pages}{072} (\bibinfo{year}{2012}), \eprint{1203.6362}.

\bibitem[{\citenamefont{Ma and Zhou}(2022)}]{Ma:2022ihn}
\bibinfo{author}{\bibfnamefont{W.-J.} \bibnamefont{Ma}} \bibnamefont{and} \bibinfo{author}{\bibfnamefont{X.}~\bibnamefont{Zhou}}, \bibinfo{journal}{JHEP} \textbf{\bibinfo{volume}{08}}, \bibinfo{pages}{107} (\bibinfo{year}{2022}), \eprint{2204.13419}.

\bibitem[{\citenamefont{Bissi et~al.}(2022)\citenamefont{Bissi, Fardelli, and Manenti}}]{Bissi:2021hjk}
\bibinfo{author}{\bibfnamefont{A.}~\bibnamefont{Bissi}}, \bibinfo{author}{\bibfnamefont{G.}~\bibnamefont{Fardelli}}, \bibnamefont{and} \bibinfo{author}{\bibfnamefont{A.}~\bibnamefont{Manenti}}, \bibinfo{journal}{JHEP} \textbf{\bibinfo{volume}{04}}, \bibinfo{pages}{016} (\bibinfo{year}{2022}), \eprint{2111.06857}.

\bibitem[{\citenamefont{Ceplak et~al.}(2021)\citenamefont{Ceplak, Giusto, Hughes, and Russo}}]{Ceplak:2021wzz}
\bibinfo{author}{\bibfnamefont{N.}~\bibnamefont{Ceplak}}, \bibinfo{author}{\bibfnamefont{S.}~\bibnamefont{Giusto}}, \bibinfo{author}{\bibfnamefont{M.~R.~R.} \bibnamefont{Hughes}}, \bibnamefont{and} \bibinfo{author}{\bibfnamefont{R.}~\bibnamefont{Russo}}, \bibinfo{journal}{JHEP} \textbf{\bibinfo{volume}{09}}, \bibinfo{pages}{204} (\bibinfo{year}{2021}), \eprint{2105.04670}.

\bibitem[{\citenamefont{Lin et~al.}(2004)\citenamefont{Lin, Lunin, and Maldacena}}]{Lin:2004nb}
\bibinfo{author}{\bibfnamefont{H.}~\bibnamefont{Lin}}, \bibinfo{author}{\bibfnamefont{O.}~\bibnamefont{Lunin}}, \bibnamefont{and} \bibinfo{author}{\bibfnamefont{J.~M.} \bibnamefont{Maldacena}}, \bibinfo{journal}{JHEP} \textbf{\bibinfo{volume}{10}}, \bibinfo{pages}{025} (\bibinfo{year}{2004}), \eprint{hep-th/0409174}.

\bibitem[{\citenamefont{Lunin and Mathur}(2002)}]{Lunin:2001jy}
\bibinfo{author}{\bibfnamefont{O.}~\bibnamefont{Lunin}} \bibnamefont{and} \bibinfo{author}{\bibfnamefont{S.~D.} \bibnamefont{Mathur}}, \bibinfo{journal}{Nucl. Phys. B} \textbf{\bibinfo{volume}{623}}, \bibinfo{pages}{342} (\bibinfo{year}{2002}), \eprint{hep-th/0109154}.

\bibitem[{\citenamefont{Maldacena and Maoz}(2002)}]{Maldacena:2000dr}
\bibinfo{author}{\bibfnamefont{J.~M.} \bibnamefont{Maldacena}} \bibnamefont{and} \bibinfo{author}{\bibfnamefont{L.}~\bibnamefont{Maoz}}, \bibinfo{journal}{JHEP} \textbf{\bibinfo{volume}{12}}, \bibinfo{pages}{055} (\bibinfo{year}{2002}), \eprint{hep-th/0012025}.

\bibitem[{\citenamefont{Kanitscheider et~al.}(2007)\citenamefont{Kanitscheider, Skenderis, and Taylor}}]{Kanitscheider:2007wq}
\bibinfo{author}{\bibfnamefont{I.}~\bibnamefont{Kanitscheider}}, \bibinfo{author}{\bibfnamefont{K.}~\bibnamefont{Skenderis}}, \bibnamefont{and} \bibinfo{author}{\bibfnamefont{M.}~\bibnamefont{Taylor}}, \bibinfo{journal}{JHEP} \textbf{\bibinfo{volume}{06}}, \bibinfo{pages}{056} (\bibinfo{year}{2007}), \eprint{0704.0690}.

\bibitem[{\citenamefont{Galliani et~al.}(2017)\citenamefont{Galliani, Giusto, and Russo}}]{Galliani:2017jlg}
\bibinfo{author}{\bibfnamefont{A.}~\bibnamefont{Galliani}}, \bibinfo{author}{\bibfnamefont{S.}~\bibnamefont{Giusto}}, \bibnamefont{and} \bibinfo{author}{\bibfnamefont{R.}~\bibnamefont{Russo}}, \bibinfo{journal}{JHEP} \textbf{\bibinfo{volume}{10}}, \bibinfo{pages}{040} (\bibinfo{year}{2017}), \eprint{1705.09250}.

\bibitem[{\citenamefont{Bombini et~al.}(2018)\citenamefont{Bombini, Galliani, Giusto, Moscato, and Russo}}]{Bombini:2017sge}
\bibinfo{author}{\bibfnamefont{A.}~\bibnamefont{Bombini}}, \bibinfo{author}{\bibfnamefont{A.}~\bibnamefont{Galliani}}, \bibinfo{author}{\bibfnamefont{S.}~\bibnamefont{Giusto}}, \bibinfo{author}{\bibfnamefont{E.}~\bibnamefont{Moscato}}, \bibnamefont{and} \bibinfo{author}{\bibfnamefont{R.}~\bibnamefont{Russo}}, \bibinfo{journal}{Eur. Phys. J. C} \textbf{\bibinfo{volume}{78}}, \bibinfo{pages}{8} (\bibinfo{year}{2018}), \eprint{1710.06820}.

\bibitem[{\citenamefont{Bombini and Galliani}(2019)}]{Bombini:2019vnc}
\bibinfo{author}{\bibfnamefont{A.}~\bibnamefont{Bombini}} \bibnamefont{and} \bibinfo{author}{\bibfnamefont{A.}~\bibnamefont{Galliani}}, \bibinfo{journal}{JHEP} \textbf{\bibinfo{volume}{06}}, \bibinfo{pages}{044} (\bibinfo{year}{2019}), \eprint{1904.02656}.

\bibitem[{\citenamefont{Giusto et~al.}(2019{\natexlab{a}})\citenamefont{Giusto, Russo, and Wen}}]{Giusto:2018ovt}
\bibinfo{author}{\bibfnamefont{S.}~\bibnamefont{Giusto}}, \bibinfo{author}{\bibfnamefont{R.}~\bibnamefont{Russo}}, \bibnamefont{and} \bibinfo{author}{\bibfnamefont{C.}~\bibnamefont{Wen}}, \bibinfo{journal}{JHEP} \textbf{\bibinfo{volume}{03}}, \bibinfo{pages}{096} (\bibinfo{year}{2019}{\natexlab{a}}), \eprint{1812.06479}.

\bibitem[{\citenamefont{Rastelli et~al.}(2019)\citenamefont{Rastelli, Roumpedakis, and Zhou}}]{Rastelli:2019gtj}
\bibinfo{author}{\bibfnamefont{L.}~\bibnamefont{Rastelli}}, \bibinfo{author}{\bibfnamefont{K.}~\bibnamefont{Roumpedakis}}, \bibnamefont{and} \bibinfo{author}{\bibfnamefont{X.}~\bibnamefont{Zhou}}, \bibinfo{journal}{JHEP} \textbf{\bibinfo{volume}{10}}, \bibinfo{pages}{140} (\bibinfo{year}{2019}), \eprint{1905.11983}.

\bibitem[{\citenamefont{Giusto et~al.}(2019{\natexlab{b}})\citenamefont{Giusto, Russo, Tyukov, and Wen}}]{Giusto:2019pxc}
\bibinfo{author}{\bibfnamefont{S.}~\bibnamefont{Giusto}}, \bibinfo{author}{\bibfnamefont{R.}~\bibnamefont{Russo}}, \bibinfo{author}{\bibfnamefont{A.}~\bibnamefont{Tyukov}}, \bibnamefont{and} \bibinfo{author}{\bibfnamefont{C.}~\bibnamefont{Wen}}, \bibinfo{journal}{JHEP} \textbf{\bibinfo{volume}{09}}, \bibinfo{pages}{030} (\bibinfo{year}{2019}{\natexlab{b}}), \eprint{1905.12314}.

\bibitem[{\citenamefont{Giusto et~al.}(2020)\citenamefont{Giusto, Russo, Tyukov, and Wen}}]{Giusto:2020neo}
\bibinfo{author}{\bibfnamefont{S.}~\bibnamefont{Giusto}}, \bibinfo{author}{\bibfnamefont{R.}~\bibnamefont{Russo}}, \bibinfo{author}{\bibfnamefont{A.}~\bibnamefont{Tyukov}}, \bibnamefont{and} \bibinfo{author}{\bibfnamefont{C.}~\bibnamefont{Wen}}, \bibinfo{journal}{Eur. Phys. J. C} \textbf{\bibinfo{volume}{80}}, \bibinfo{pages}{736} (\bibinfo{year}{2020}), \eprint{2005.08560}.


\bibitem[{\citenamefont{Liu et~al.}(2007)\citenamefont{Liu, Lu, Pope, and Vazquez-Poritz}}]{Liu:2007xj}
\bibinfo{author}{\bibfnamefont{J.~T.} \bibnamefont{Liu}}, \bibinfo{author}{\bibfnamefont{H.}~\bibnamefont{Lu}}, \bibinfo{author}{\bibfnamefont{C.~N.} \bibnamefont{Pope}}, \bibnamefont{and} \bibinfo{author}{\bibfnamefont{J.~F.} \bibnamefont{Vazquez-Poritz}}, \bibinfo{journal}{JHEP} \textbf{\bibinfo{volume}{10}}, \bibinfo{pages}{030} (\bibinfo{year}{2007}), \eprint{hep-th/0703184}.

\bibitem[{\citenamefont{Giusto and Rosso}(2024)}]{Giusto:2024trt}
\bibinfo{author}{\bibfnamefont{S.}~\bibnamefont{Giusto}} \bibnamefont{and} \bibinfo{author}{\bibfnamefont{A.}~\bibnamefont{Rosso}} (\bibinfo{year}{2024}), \eprint{2401.01254}.

\bibitem[{\citenamefont{Skenderis and Taylor}(2006)}]{Skenderis:2006uy}
\bibinfo{author}{\bibfnamefont{K.}~\bibnamefont{Skenderis}} \bibnamefont{and} \bibinfo{author}{\bibfnamefont{M.}~\bibnamefont{Taylor}}, \bibinfo{journal}{JHEP} \textbf{\bibinfo{volume}{05}}, \bibinfo{pages}{057} (\bibinfo{year}{2006}), \eprint{hep-th/0603016}.

\bibitem[{\citenamefont{Skenderis and Taylor}(2007)}]{Skenderis:2007yb}
\bibinfo{author}{\bibfnamefont{K.}~\bibnamefont{Skenderis}} \bibnamefont{and} \bibinfo{author}{\bibfnamefont{M.}~\bibnamefont{Taylor}}, \bibinfo{journal}{JHEP} \textbf{\bibinfo{volume}{09}}, \bibinfo{pages}{019} (\bibinfo{year}{2007}), \eprint{0706.0216}.

\bibitem[{\citenamefont{Ganchev et~al.}(2023)\citenamefont{Ganchev, Giusto, Houppe, Russo, and Warner}}]{Ganchev:2023sth}
\bibinfo{author}{\bibfnamefont{B.}~\bibnamefont{Ganchev}}, \bibinfo{author}{\bibfnamefont{S.}~\bibnamefont{Giusto}}, \bibinfo{author}{\bibfnamefont{A.}~\bibnamefont{Houppe}}, \bibinfo{author}{\bibfnamefont{R.}~\bibnamefont{Russo}}, \bibnamefont{and} \bibinfo{author}{\bibfnamefont{N.~P.} \bibnamefont{Warner}}, \bibinfo{journal}{JHEP} \textbf{\bibinfo{volume}{10}}, \bibinfo{pages}{163} (\bibinfo{year}{2023}), \eprint{2307.13021}.

\bibitem[{\citenamefont{Usyukina and Davydychev}(1993{\natexlab{a}})}]{Usyukina:1992jd}
\bibinfo{author}{\bibfnamefont{N.~I.} \bibnamefont{Usyukina}} \bibnamefont{and} \bibinfo{author}{\bibfnamefont{A.~I.} \bibnamefont{Davydychev}}, \bibinfo{journal}{Phys. Lett. B} \textbf{\bibinfo{volume}{298}}, \bibinfo{pages}{363} (\bibinfo{year}{1993}{\natexlab{a}}).

\bibitem[{\citenamefont{Usyukina and Davydychev}(1993{\natexlab{b}})}]{Usyukina:1993ch}
\bibinfo{author}{\bibfnamefont{N.~I.} \bibnamefont{Usyukina}} \bibnamefont{and} \bibinfo{author}{\bibfnamefont{A.~I.} \bibnamefont{Davydychev}}, \bibinfo{journal}{Phys. Lett. B} \textbf{\bibinfo{volume}{305}}, \bibinfo{pages}{136} (\bibinfo{year}{1993}{\natexlab{b}}).

\bibitem[{\citenamefont{Drummond et~al.}(2007)\citenamefont{Drummond, Gallot, and Sokatchev}}]{Drummond:2006by}
\bibinfo{author}{\bibfnamefont{J.~M.} \bibnamefont{Drummond}}, \bibinfo{author}{\bibfnamefont{L.}~\bibnamefont{Gallot}}, \bibnamefont{and} \bibinfo{author}{\bibfnamefont{E.}~\bibnamefont{Sokatchev}}, \bibinfo{journal}{Phys. Lett. B} \textbf{\bibinfo{volume}{645}}, \bibinfo{pages}{95} (\bibinfo{year}{2007}), \eprint{hep-th/0610280}.

\bibitem[{\citenamefont{Gon\c{c}alves}(2015)}]{Goncalves:2014ffa}
\bibinfo{author}{\bibfnamefont{V.}~\bibnamefont{Gon\c{c}alves}}, \bibinfo{journal}{JHEP} \textbf{\bibinfo{volume}{04}}, \bibinfo{pages}{150} (\bibinfo{year}{2015}), \eprint{1411.1675}.

\bibitem[{\citenamefont{Paul et~al.}(2023)\citenamefont{Paul, Perlmutter, and Raj}}]{Paul:2023rka}
\bibinfo{author}{\bibfnamefont{H.}~\bibnamefont{Paul}}, \bibinfo{author}{\bibfnamefont{E.}~\bibnamefont{Perlmutter}}, \bibnamefont{and} \bibinfo{author}{\bibfnamefont{H.}~\bibnamefont{Raj}}, \bibinfo{journal}{JHEP} \textbf{\bibinfo{volume}{08}}, \bibinfo{pages}{078} (\bibinfo{year}{2023}), \eprint{2303.13207}.

\bibitem[{\citenamefont{Brown et~al.}(2023)\citenamefont{Brown, Wen, and Xie}}]{Brown:2023why}
\bibinfo{author}{\bibfnamefont{A.}~\bibnamefont{Brown}}, \bibinfo{author}{\bibfnamefont{C.}~\bibnamefont{Wen}}, \bibnamefont{and} \bibinfo{author}{\bibfnamefont{H.}~\bibnamefont{Xie}}, \bibinfo{journal}{JHEP} \textbf{\bibinfo{volume}{07}}, \bibinfo{pages}{129} (\bibinfo{year}{2023}), \eprint{2303.17570}.

\bibitem[{\citenamefont{Caetano et~al.}(2024)\citenamefont{Caetano, Komatsu, and Wang}}]{Caetano:2023zwe}
\bibinfo{author}{\bibfnamefont{J.~a.} \bibnamefont{Caetano}}, \bibinfo{author}{\bibfnamefont{S.}~\bibnamefont{Komatsu}}, \bibnamefont{and} \bibinfo{author}{\bibfnamefont{Y.}~\bibnamefont{Wang}}, \bibinfo{journal}{JHEP} \textbf{\bibinfo{volume}{02}}, \bibinfo{pages}{047} (\bibinfo{year}{2024}), \eprint{2306.00929}.

\bibitem[{\citenamefont{Brown et~al.}(2024)\citenamefont{Brown, Galvagno, and Wen}}]{Brown:2024yvt}
\bibinfo{author}{\bibfnamefont{A.}~\bibnamefont{Brown}}, \bibinfo{author}{\bibfnamefont{F.}~\bibnamefont{Galvagno}}, \bibnamefont{and} \bibinfo{author}{\bibfnamefont{C.}~\bibnamefont{Wen}} (\bibinfo{year}{2024}), \eprint{2407.02250}.

\bibitem[{\citenamefont{Dolan and Osborn}(2006)}]{Dolan:2004iy}
\bibinfo{author}{\bibfnamefont{F.~A.} \bibnamefont{Dolan}} \bibnamefont{and} \bibinfo{author}{\bibfnamefont{H.}~\bibnamefont{Osborn}}, \bibinfo{journal}{Annals Phys.} \textbf{\bibinfo{volume}{321}}, \bibinfo{pages}{581} (\bibinfo{year}{2006}), \eprint{hep-th/0412335}.

\bibitem[{\citenamefont{Turton and Tyukov}(2024)}]{Turton:2024afd}
\bibinfo{author}{\bibfnamefont{D.}~\bibnamefont{Turton}} \bibnamefont{and} \bibinfo{author}{\bibfnamefont{A.}~\bibnamefont{Tyukov}} (\bibinfo{year}{2024}), \eprint{2408.16834}.

\bibitem[{\citenamefont{Mack}(2009)}]{Mack:2009mi}
\bibinfo{author}{\bibfnamefont{G.}~\bibnamefont{Mack}} (\bibinfo{year}{2009}), \eprint{0907.2407}.

\bibitem[{\citenamefont{Penedones}(2011)}]{Penedones:2010ue}
\bibinfo{author}{\bibfnamefont{J.}~\bibnamefont{Penedones}}, \bibinfo{journal}{JHEP} \textbf{\bibinfo{volume}{03}}, \bibinfo{pages}{025} (\bibinfo{year}{2011}), \eprint{1011.1485}.

\bibitem[{\citenamefont{Fitzpatrick et~al.}(2011)\citenamefont{Fitzpatrick, Kaplan, Penedones, Raju, and van Rees}}]{Fitzpatrick:2011ia}
\bibinfo{author}{\bibfnamefont{A.~L.} \bibnamefont{Fitzpatrick}}, \bibinfo{author}{\bibfnamefont{J.}~\bibnamefont{Kaplan}}, \bibinfo{author}{\bibfnamefont{J.}~\bibnamefont{Penedones}}, \bibinfo{author}{\bibfnamefont{S.}~\bibnamefont{Raju}}, \bibnamefont{and} \bibinfo{author}{\bibfnamefont{B.~C.} \bibnamefont{van Rees}}, \bibinfo{journal}{JHEP} \textbf{\bibinfo{volume}{11}}, \bibinfo{pages}{095} (\bibinfo{year}{2011}), \eprint{1107.1499}.

\bibitem[{\citenamefont{Rastelli and Zhou}(2017)}]{Rastelli:2016nze}
\bibinfo{author}{\bibfnamefont{L.}~\bibnamefont{Rastelli}} \bibnamefont{and} \bibinfo{author}{\bibfnamefont{X.}~\bibnamefont{Zhou}}, \bibinfo{journal}{Phys. Rev. Lett.} \textbf{\bibinfo{volume}{118}}, \bibinfo{pages}{091602} (\bibinfo{year}{2017}), \eprint{1608.06624}.

\bibitem[{\citenamefont{Aprile and Vieira}(2020)}]{Aprile:2020luw}
\bibinfo{author}{\bibfnamefont{F.}~\bibnamefont{Aprile}} \bibnamefont{and} \bibinfo{author}{\bibfnamefont{P.}~\bibnamefont{Vieira}}, \bibinfo{journal}{JHEP} \textbf{\bibinfo{volume}{12}}, \bibinfo{pages}{206} (\bibinfo{year}{2020}), \eprint{2007.09176}.

\bibitem[{\citenamefont{Allendes et~al.}(2013)\citenamefont{Allendes, Kniehl, Kondrashuk, Cuello, and Medar}}]{Allendes:2012mr}
\bibinfo{author}{\bibfnamefont{P.}~\bibnamefont{Allendes}}, \bibinfo{author}{\bibfnamefont{B.}~\bibnamefont{Kniehl}}, \bibinfo{author}{\bibfnamefont{I.}~\bibnamefont{Kondrashuk}}, \bibinfo{author}{\bibfnamefont{E.~A.~N.} \bibnamefont{Cuello}}, \bibnamefont{and} \bibinfo{author}{\bibfnamefont{M.~R.} \bibnamefont{Medar}}, \bibinfo{journal}{Nucl. Phys. B} \textbf{\bibinfo{volume}{870}}, \bibinfo{pages}{243} (\bibinfo{year}{2013}), \eprint{1205.6257}.

\bibitem[{\citenamefont{Alday and Bissi}(2017)}]{Alday:2017xua}
\bibinfo{author}{\bibfnamefont{L.~F.} \bibnamefont{Alday}} \bibnamefont{and} \bibinfo{author}{\bibfnamefont{A.}~\bibnamefont{Bissi}}, \bibinfo{journal}{Phys. Rev. Lett.} \textbf{\bibinfo{volume}{119}}, \bibinfo{pages}{171601} (\bibinfo{year}{2017}), \eprint{1706.02388}.

\bibitem[{\citenamefont{Aprile et~al.}(2018{\natexlab{a}})\citenamefont{Aprile, Drummond, Heslop, and Paul}}]{Aprile:2017bgs}
\bibinfo{author}{\bibfnamefont{F.}~\bibnamefont{Aprile}}, \bibinfo{author}{\bibfnamefont{J.~M.} \bibnamefont{Drummond}}, \bibinfo{author}{\bibfnamefont{P.}~\bibnamefont{Heslop}}, \bibnamefont{and} \bibinfo{author}{\bibfnamefont{H.}~\bibnamefont{Paul}}, \bibinfo{journal}{JHEP} \textbf{\bibinfo{volume}{01}}, \bibinfo{pages}{035} (\bibinfo{year}{2018}{\natexlab{a}}), \eprint{1706.02822}.

\bibitem[{\citenamefont{Alday and Caron-Huot}(2018)}]{Alday:2017vkk}
\bibinfo{author}{\bibfnamefont{L.~F.} \bibnamefont{Alday}} \bibnamefont{and} \bibinfo{author}{\bibfnamefont{S.}~\bibnamefont{Caron-Huot}}, \bibinfo{journal}{JHEP} \textbf{\bibinfo{volume}{12}}, \bibinfo{pages}{017} (\bibinfo{year}{2018}), \eprint{1711.02031}.

\bibitem[{\citenamefont{Alday}(2021)}]{Alday:2018kkw}
\bibinfo{author}{\bibfnamefont{L.~F.} \bibnamefont{Alday}}, \bibinfo{journal}{JHEP} \textbf{\bibinfo{volume}{04}}, \bibinfo{pages}{005} (\bibinfo{year}{2021}), \eprint{1812.11783}.

\bibitem[{\citenamefont{Aprile et~al.}(2020{\natexlab{b}})\citenamefont{Aprile, Drummond, Heslop, and Paul}}]{Aprile:2019rep}
\bibinfo{author}{\bibfnamefont{F.}~\bibnamefont{Aprile}}, \bibinfo{author}{\bibfnamefont{J.}~\bibnamefont{Drummond}}, \bibinfo{author}{\bibfnamefont{P.}~\bibnamefont{Heslop}}, \bibnamefont{and} \bibinfo{author}{\bibfnamefont{H.}~\bibnamefont{Paul}}, \bibinfo{journal}{JHEP} \textbf{\bibinfo{volume}{03}}, \bibinfo{pages}{190} (\bibinfo{year}{2020}{\natexlab{b}}), \eprint{1912.01047}.

\bibitem[{\citenamefont{Doobary and Heslop}(2015)}]{Doobary:2015gia}
\bibinfo{author}{\bibfnamefont{R.}~\bibnamefont{Doobary}} \bibnamefont{and} \bibinfo{author}{\bibfnamefont{P.}~\bibnamefont{Heslop}}, \bibinfo{journal}{JHEP} \textbf{\bibinfo{volume}{12}}, \bibinfo{pages}{159} (\bibinfo{year}{2015}), \eprint{1508.03611}.

\bibitem[{\citenamefont{Aprile et~al.}(2021)\citenamefont{Aprile, Drummond, Paul, and Santagata}}]{Aprile:2020mus}
\bibinfo{author}{\bibfnamefont{F.}~\bibnamefont{Aprile}}, \bibinfo{author}{\bibfnamefont{J.~M.} \bibnamefont{Drummond}}, \bibinfo{author}{\bibfnamefont{H.}~\bibnamefont{Paul}}, \bibnamefont{and} \bibinfo{author}{\bibfnamefont{M.}~\bibnamefont{Santagata}}, \bibinfo{journal}{JHEP} \textbf{\bibinfo{volume}{11}}, \bibinfo{pages}{109} (\bibinfo{year}{2021}), \eprint{2012.12092}.

\bibitem[{\citenamefont{Huang et~al.}(2024{\natexlab{a}})\citenamefont{Huang, Wang, and Yuan}}]{Huang:2024dck}
\bibinfo{author}{\bibfnamefont{Z.}~\bibnamefont{Huang}}, \bibinfo{author}{\bibfnamefont{B.}~\bibnamefont{Wang}}, \bibnamefont{and} \bibinfo{author}{\bibfnamefont{E.~Y.} \bibnamefont{Yuan}}, \bibinfo{journal}{JHEP} \textbf{\bibinfo{volume}{07}}, \bibinfo{pages}{176} (\bibinfo{year}{2024}{\natexlab{a}}), \eprint{2403.10607}.

\bibitem[{\citenamefont{Huang et~al.}(2024{\natexlab{b}})\citenamefont{Huang, Wang, and Yuan}}]{Huang:2024rxr}
\bibinfo{author}{\bibfnamefont{Z.}~\bibnamefont{Huang}}, \bibinfo{author}{\bibfnamefont{B.}~\bibnamefont{Wang}}, \bibnamefont{and} \bibinfo{author}{\bibfnamefont{E.~Y.} \bibnamefont{Yuan}} (\bibinfo{year}{2024}{\natexlab{b}}), \eprint{2407.03408}.

\bibitem[{\citenamefont{Gon\c{c}alves et~al.}(2019)\citenamefont{Gon\c{c}alves, Pereira, and Zhou}}]{Goncalves:2019znr}
\bibinfo{author}{\bibfnamefont{V.}~\bibnamefont{Gon\c{c}alves}}, \bibinfo{author}{\bibfnamefont{R.}~\bibnamefont{Pereira}}, \bibnamefont{and} \bibinfo{author}{\bibfnamefont{X.}~\bibnamefont{Zhou}}, \bibinfo{journal}{JHEP} \textbf{\bibinfo{volume}{10}}, \bibinfo{pages}{247} (\bibinfo{year}{2019}), \eprint{1906.05305}.

\bibitem[{\citenamefont{Gon\c{c}alves et~al.}(2023)\citenamefont{Gon\c{c}alves, Meneghelli, Pereira, Vilas~Boas, and Zhou}}]{Goncalves:2023oyx}
\bibinfo{author}{\bibfnamefont{V.}~\bibnamefont{Gon\c{c}alves}}, \bibinfo{author}{\bibfnamefont{C.}~\bibnamefont{Meneghelli}}, \bibinfo{author}{\bibfnamefont{R.}~\bibnamefont{Pereira}}, \bibinfo{author}{\bibfnamefont{J.}~\bibnamefont{Vilas~Boas}}, \bibnamefont{and} \bibinfo{author}{\bibfnamefont{X.}~\bibnamefont{Zhou}}, \bibinfo{journal}{JHEP} \textbf{\bibinfo{volume}{08}}, \bibinfo{pages}{067} (\bibinfo{year}{2023}), \eprint{2302.01896}.

\bibitem[{\citenamefont{Alday et~al.}(2024)\citenamefont{Alday, Gon\c{c}alves, Nocchi, and Zhou}}]{Alday:2023kfm}
\bibinfo{author}{\bibfnamefont{L.~F.} \bibnamefont{Alday}}, \bibinfo{author}{\bibfnamefont{V.}~\bibnamefont{Gon\c{c}alves}}, \bibinfo{author}{\bibfnamefont{M.}~\bibnamefont{Nocchi}}, \bibnamefont{and} \bibinfo{author}{\bibfnamefont{X.}~\bibnamefont{Zhou}}, \bibinfo{journal}{Phys. Rev. Res.} \textbf{\bibinfo{volume}{6}}, \bibinfo{pages}{L012041} (\bibinfo{year}{2024}), \eprint{2307.06884}.

\bibitem[{\citenamefont{Cao et~al.}(2024)\citenamefont{Cao, He, and Tang}}]{Cao:2023cwa}
\bibinfo{author}{\bibfnamefont{Q.}~\bibnamefont{Cao}}, \bibinfo{author}{\bibfnamefont{S.}~\bibnamefont{He}}, \bibnamefont{and} \bibinfo{author}{\bibfnamefont{Y.}~\bibnamefont{Tang}}, \bibinfo{journal}{Phys. Rev. Lett.} \textbf{\bibinfo{volume}{133}}, \bibinfo{pages}{021605} (\bibinfo{year}{2024}), \eprint{2312.15484}.

\bibitem[{\citenamefont{Aprile et~al.}(2018{\natexlab{b}})\citenamefont{Aprile, Drummond, Heslop, and Paul}}]{Aprile:2017xsp}
\bibinfo{author}{\bibfnamefont{F.}~\bibnamefont{Aprile}}, \bibinfo{author}{\bibfnamefont{J.~M.} \bibnamefont{Drummond}}, \bibinfo{author}{\bibfnamefont{P.}~\bibnamefont{Heslop}}, \bibnamefont{and} \bibinfo{author}{\bibfnamefont{H.}~\bibnamefont{Paul}}, \bibinfo{journal}{JHEP} \textbf{\bibinfo{volume}{02}}, \bibinfo{pages}{133} (\bibinfo{year}{2018}{\natexlab{b}}), \eprint{1706.08456}.

\bibitem[{\citenamefont{Aharony et~al.}(2008)\citenamefont{Aharony, Bergman, Jafferis, and Maldacena}}]{Aharony:2008ug}
\bibinfo{author}{\bibfnamefont{O.}~\bibnamefont{Aharony}}, \bibinfo{author}{\bibfnamefont{O.}~\bibnamefont{Bergman}}, \bibinfo{author}{\bibfnamefont{D.~L.} \bibnamefont{Jafferis}}, \bibnamefont{and} \bibinfo{author}{\bibfnamefont{J.}~\bibnamefont{Maldacena}}, \bibinfo{journal}{JHEP} \textbf{\bibinfo{volume}{10}}, \bibinfo{pages}{091} (\bibinfo{year}{2008}), \eprint{0806.1218}.

\bibitem[{\citenamefont{Chester}(2020)}]{Chester:2019pvm}
\bibinfo{author}{\bibfnamefont{S.~M.} \bibnamefont{Chester}}, \bibinfo{journal}{JHEP} \textbf{\bibinfo{volume}{04}}, \bibinfo{pages}{193} (\bibinfo{year}{2020}), \eprint{1908.05247}.


\bibitem{Bissi:2024tqf}
A.~Bissi, G.~Fardelli and A.~Manenti, 
JHEP \textbf{07}, 074 (2025), 
arXiv:2412.19788.

\bibitem{Aprile:2025hlt}
F.~Aprile, S.~Giusto and R.~Russo, JHEP (2025)
arXiv:2503.02855

\end{thebibliography}
\end{document}